\documentclass[aps,pra,floatfix,twocolumn,showpacs]{revtex4}
\usepackage{epsfig}
\usepackage{graphicx}
\usepackage{dcolumn}
\begin{document}
\title{On the use of the $V^{N-M}$ approximation in atomic calculations.}
\author{V. A. Dzuba}
\email{V.Dzuba@unsw.edu.au}
\affiliation{School of Physics, University of New South Wales, Sydney 2052,
Australia}

\date{\today}

\begin{abstract}

We demonstrate that $V^{N-M}$ approximation is a good starting point
for the configuration interaction calculations for many-electron
atoms and ions. $N$ is the total number of electrons in the neutral
atom, $M$ is the number of valence electrons. $V^{N-M}$ is the
self-consistent Hartree Fock potential for a closed-shell ion with
all valence electrons removed. Using of the $V^{N-M}$ approximation
considerably simplifies the many-body perturbation theory for the 
core-valence correlations. It makes it easier to include higher-order
correlations which often significantly improves the accuracy of the
calculations. Calculations for krypton and barium and their positive 
ions are presented for illustration.

\end{abstract}
\pacs{PACS: 31.25.-v, 31.25.Eb, 31.25.Jf}

\maketitle

\section{introduction}

Atomic physics is a valuable tool to study many fundamental problems.
It is used to study parity and time invariance violating
interactions (see, e.g. \cite{ginges}),
possible variation of the fundamental constants in
quasar absorption spectra \cite{Webb} and in present-day experiments
by comparing the rates of different atomic clocks \cite{flambaum}, etc.
However, interpretation of the atomic measurements is often limited
by accuracy of atomic calculations. For example, the accuracy of the
most precise measurements of atomic parity non-conserving effects (PNC) 
in atoms which has been achieved for cesium is 0.35\% \cite{Wood}. 
The accuracy of the best calculations is from 
0.5\% to 1\% \cite{Dzuba02,Dzuba89,Blundell90,Blundell92}. 
Situation is even worse for thallium. Experimental accuracy of the PNC
measurements is 1\% \cite{Fortson} while best theoretical accuracy is from 
2.5\% \cite{Kozlov} to 3\% \cite{DzubaTl}. On this level of accuracy there
is perfect agreement of the PNC measurements with the standard model
and any further progress would need significant improvement in atomic
theory. There are many other examples where accurate atomic calculations 
are needed. These include atomic clocks, quantum computations, plasma 
physics, etc. Therefore, it is worth to study the ways of improving
the methods of calculations.

It is well known that the perturbation theory in residual Coulomb
interaction converge very poorly for many electron atoms and some
all-order technique is needed to achieve good accuracy of calculations.
For atoms with one external electron above closed shells there are at
least two all-order methods which lead to a fraction of percent accuracy
in calculation of the energies as compared to experimental data. One
is an all-order correlation potential method (also called perturbation
theory in screened Coulomb interaction) \cite{Dzuba88}. Another is
linearized coupled cluster approach (CC)\cite{Liu89}. For atoms with
more than one external electron good accuracy can be achieved when
different methods are combined to include correlations between
valence electrons together with the core-valence correlations. 
This can be done by combining configuration interaction method
with the many-body perturbation theory (CI+MBPT) \cite{Kozlov96}
or CC method with the MBPT \cite{Johnson91} or with the CI method
\cite{Kozlov04}.

The key question in developing of all these methods is where to start
or what potential to chose to generate a complete set of single-electron
states. It is well accepted now that the Hartree Fock potential is the best
choice for the perturbation theory expansion. This is because self-consistency
condition leads to exact cancellation between Coulomb and potential terms
in the residual interaction so that potential terms are completely eliminated
from the perturbation theory expansion. The natural choice for atoms with one
external electron is the $V^{N-1}$ Hartree Fock potential introduced by
Kelly \cite{Kelly}. In the $V^{N-1}$ approximation the self-consistency
procedure is initially done for a closed-shell positive ion. States of
external electron are then calculated in the field of frozen core.
There is exact cancellation between direct and exchange self-action terms 
in the Hartree Fock potential for closed shell systems. 
Therefore, by including self-action, we can easily see that
states in the core and states above core are calculated in the same
potential. Other words, $V^{N-1}$ potential generates a complete set of 
orthogonal single-electron states which are convenient for use in the
perturbation theory expansion. Using this set in an appropriate
all-oder method leads to very good results for a neutral atom in spite
of the fact that the core of the atom is actually the core of a
positive ion.

The $V^{N-1}$ approximation can also be used for atoms with more than one
external electron. However, in this case the system of $N-1$ electrons
is most likely to be an open-shell system and some averaging procedure
is needed to define the $V^{N-1}$ potential. Another complication
arise when core-valence correlation are to be included by means
of MBPT. There is no exact cancellation between potential terms any more.
The potential in the effective Hamiltonian is now $V^{N-M}$ potential,
where $M$ is number of valence electrons and $M>1$. Perturbation theory
expansion would have terms proportional to $V^{N-M} - V^{N-1}$.
These terms are called {\em subtraction diagrams} \cite{Kozlov96} or
{\em $\Delta$-terms} \cite{Johnson}. The number of these terms is larger
than number of pure Coulomb terms and this represents significant
complication of the MBPT. These terms can be totally avoided if calculations
from the very beginning are done in the $V^{N-M}$ potential. However,
it is widely believed that doing calculations for a neutral atom
by starting from a highly charged ion would lead to poor convergence
of the perturbation expansion and poor end results. Indeed, after
the initial Hartree Fock procedure is done  the core is kept frozen
in all consequent calculations. No further perturbation expansion can
change anything in the core, leaving it to be the core of the highly
charged ion.

The purpose of this work is to demonstrate that the core of the 
highly charged ion is often not very much different from the core
of neutral atom and $V^{N-M}$ approximation can be a good approximation
for atoms with several valence electrons. The main gain is total
elimination of subtraction diagrams. This significantly
simplifies the  perturbation theory expansion for the core-valence 
correlations.
It is also much easier to include higher-order core-valence correlations
in the $V^{N-M}$ approximation. Inclusion of higher-oder
correlations can significantly improve the accuracy of the calculations.

We consider CI+MBPT calculations for neutral krypton and barium and
their positive ions to illustrate the advantage of the $V^{N-M}$
approximation.

\section{Calculations}

\subsection{Krypton.}

Let's start our consideration from an extreme case - an atom with eight
valence electrons. The purpose of this example is to illustrate that even
removal of as many as eight electrons do not lead to any dramatic
changes in the atomic core and $V^{N-8}$ approximation is still
reasonably good approximation for the neutral atom as well as for the
all chain of positive ions starting from number of valence electrons $M=1$
and up to $M=8$.

Table~\ref{tab1} compares core states of Kr~I and Kr~IX. Calculations 
are done in $V^N$ and $V^{N-8}$ potentials respectively. We present
singe-electron energies, overage radius ($\langle r \rangle$),
square root of overage square radius ($\langle r^2 \rangle^{1/2}$), position of
the maximum of the wave function ($r(f_{max}$)), the value in the maximum ($f_{max}$)
as well as the range of distances where 80\% of the electron density is
located (from $r_1$ to $r_2$). 
It is easy to see that changing from $V^N$ to $V^{N-8}$ potential has
large effect on the energies of core states but not on their wave functions.
Indeed, the energy of $3d$ states change almost two times while overage
radius (or square root of overage square radius) changes by about 2-3\% only,
position of the maximum does not change at all and the value of the wave function
in the maximum changes by about 1\% only.

\begin{table*}
\caption{\label{tab1}Parameters of core states of Kr~I and Kr~IX (atomic units).}
\begin{ruledtabular}
\begin{tabular}{lrcccccc}
State       & Energy     & $\langle r \rangle$ 
                                   & $\langle r^2 \rangle^{1/2}$ 
                                             & $r(f_{max})$ 
                                                       & $f_{max}$ & $r_1$ & $r_2$ \\
\hline
\multicolumn{8}{c}{Kr~I} \\
 $1s$       & -529.6849  & 0.0415  & 0.0481  & 0.0269  & 4.3707  & 0.0151  & 0.0731 \\
 $2s$       &  -72.0798  & 0.1827  & 0.1986  & 0.1541  & 2.4630  & 0.0987  & 0.2839 \\
 $2p_{1/2}$ &  -64.8748  & 0.1574  & 0.1744  & 0.1216  & 2.4476  & 0.0731  & 0.2605 \\
 $2p_{3/2}$ &  -62.8792  & 0.1613  & 0.1784  & 0.1253  & 2.4283  & 0.0753  & 0.2605 \\
 $3s$       &  -11.2245  & 0.5271  & 0.5648  & 0.4704  & 1.5508  & 0.3182  & 0.7794 \\
 $3p_{1/2}$ &   -8.6199  & 0.5314  & 0.5744  & 0.4577  & 1.4924  & 0.3006  & 0.7996 \\
 $3p_{3/2}$ &   -8.3128  & 0.5412  & 0.5848  & 0.4704  & 1.4800  & 0.3093  & 0.8202 \\
 $3d_{3/2}$ &   -3.7776  & 0.5505  & 0.6095  & 0.4098  & 1.3459  & 0.2681  & 0.9072 \\
 $3d_{5/2}$ &   -3.7268  & 0.5543  & 0.6136  & 0.4098  & 1.3415  & 0.2681  & 0.9072 \\
	   	     		  	    	      	        	  	        
 $4s$       &   -1.1877  & 1.6008  & 1.7136  & 1.3629  & 0.8954  & 0.9535  & 2.4031 \\
 $4p_{1/2}$ &   -0.5415  & 1.9147  & 2.0711  & 1.5253  & 0.7921  & 1.1037  & 2.9420 \\
 $4p_{3/2}$ &   -0.5143  & 1.9586  & 2.1196  & 1.5594  & 0.7825  & 1.1037  & 2.9942 \\

\multicolumn{8}{c}{Kr~IX} \\
 $1s$       &   -534.8482  & 0.0415  & 0.0481  & 0.0269  & 4.3708  & 0.0151  & 0.0731 \\
 $2s$       &    -77.1131  & 0.1827  & 0.1985  & 0.1541  & 2.4633  & 0.0987  & 0.2839 \\
 $2p_{1/2}$ &    -69.9296  & 0.1573  & 0.1743  & 0.1216  & 2.4480  & 0.0731  & 0.2605 \\
 $2p_{3/2}$ &    -67.9321  & 0.1613  & 0.1783  & 0.1253  & 2.4288  & 0.0753  & 0.2605 \\
 $3s$       &    -16.1190  & 0.5258  & 0.5630  & 0.4704  & 1.5530  & 0.3182  & 0.7794 \\
 $3p_{1/2}$ &    -13.5239  & 0.5285  & 0.5706  & 0.4577  & 1.4970  & 0.3006  & 0.7996 \\
 $3p_{3/2}$ &    -13.2140  & 0.5378  & 0.5805  & 0.4704  & 1.4851  & 0.3093  & 0.8202 \\
 $3d_{3/2}$ &     -8.6967  & 0.5376  & 0.5918  & 0.4098  & 1.3624  & 0.2605  & 0.8628 \\
 $3d_{5/2}$ &     -8.6450  & 0.5411  & 0.5955  & 0.4098  & 1.3584  & 0.2681  & 0.8848 \\
\end{tabular}
\end{ruledtabular}
\end{table*}

To understand this behavior one should look at the distances where electrons
are localized. As can be seen from Table~\ref{tab1} valence electrons ($4s$ and $4p$)
are localized at significantly larger distances than core electrons. There is almost
no overlap between densities of core and valence electrons. Indeed, 90\% of the density
of the $4s$ and $4p$ electrons are at distances $r>a_B$ ($0.95a_b$ for the $4s$
state and $1.1a_B$ for the $4p$ state) while 90\% of the density of the uppermost 
core state $3d$ is at $r<0.907a_B$. This means that valence states can only create
constant field inside the core. For example
\[     Y_{0(4s)}(r) = \int \frac{|\psi_{4s}(r')|^2}{r_>}dr' 
       \approx Const \ \ \rm{at} \ \ r<a_B.
\]
Correction to the energy of a core state is given by diagonal matrix element
\[  \delta \epsilon_n \sim \int |\psi_n(r)|^2 Y_0(r)dr. \]
This matrix element is large.

In contrast, correction to wave function is given by off-diagonal matrix elements.
These matrix elements are small due to orthogonality of wave functions:
\[    \int \psi_n(r)^{\dagger}\psi_m(r) Y_0(r)dr \approx 
      Const \int \psi_n(r)^{\dagger}\psi_m(r)dr = 0. \]

\begin{figure}
\centering
\epsfig{figure=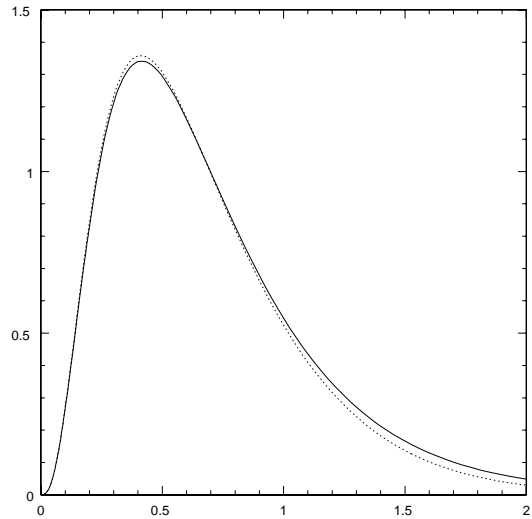,scale=0.38}
\caption{Radial wave function of the $3d_{5/2}$ state of Kr~I (solid line)
and Kr~IX (dotted line).}
\label{f3d}
\end{figure}

Fig.~\ref{f3d} shows the $3d_{5/2}$ radial wave functions of Kr~I and Kr~IX.
One can see that they are almost identical. There is some difference at large 
distances due to different energies ($\psi \sim \exp(-\sqrt{2|\epsilon|}r)$).
This difference has some effect on the normalization of the wave function
leading to small difference in the maximum. Apart from this the wave functions 
are very similar.

We see that the removal of eight valence electrons from Kr~I affects only energies
of the core states but not their wave functions. Obviously, change in the energies
affects the MBPT for the core-valence correlations through the change in energy
denominators. But what is more important is the absence of the subtraction
diagrams which makes the MBPT to be much more simple. Excitation energies
are larger in Kr~IX than in Kr~I which means that MBPT terms are smaller and
convergence is likely to be better. Therefore, it is natural to assume that
the $V^{N-8}$ approximation is a good initial approximation for all krypton
ions starting from Kr~IX and up to neutral Kr~I, with number of valence electrons 
ranges from none to eight. We have performed the calculations to check this.

Hartree Fock energy of the $3d_{5/2}$ state of Kr~IX 
(8.645 a.u., see Table~\ref{tab1}) agrees within 2\% with the experimental 
ionization energy of Kr~IX (8.488 a.u., \cite{nist}). The difference should be
mostly attributed to the correlations.

We can do much better calculations for Kr~VIII. It has one valence electron
above closed shells. We calculate its states in the field of frozen core
($V^{N-8}$ potential) in Hartree Fock and Brueckner approximations.
The latter means that we modify the HF equations for valence electron by
including correlation potential $\hat \Sigma$ (see \cite{CPM} for details).
We calculate $\hat \Sigma$ in second order of MBPT. The results are presented
in Table~\ref{kr8}. As can be seen Hartree Fock energies differ from experiment
by about 1\% while inclusion of correlations improves them significantly
brining the agreement to better than 0.1\%.

\begin{table}
\caption{\label{kr8}Energy levels of Kr~VIII (cm$^{-1}$).}
\begin{ruledtabular}
\begin{tabular}{lrrr}
\multicolumn{1}{c}{State} & HF & Brueckner & Expt\footnotemark[1] \\
\hline
  $4s$       &   1004870  &     1015504  &  1014665 \\
  $4p_{1/2}$ &    862612  &      871429  &   870970 \\
  $4p_{3/2}$ &    852990  &      861472  &   861189 \\
  $4d_{3/2}$ &    635048  &      640449  &   640618 \\
  $4d_{5/2}$ &    633695  &      639065  &   639284 \\
\end{tabular}
\end{ruledtabular}
\noindent \footnotetext[1]{NIST, \cite{nist}.}
\end{table}

We use the combined CI+MBPT method for ions with more than one valence 
electron \cite{Kozlov96}. Like in standard CI method the Schr\"{o}dinger
equation is written for the many-electron wave function of valence
electrons
\begin{equation}
  (\hat H^{\rm eff} - E)\Psi = 0.
\label{Schr}
\end{equation}
$\Psi$ has a form of expansion over single-determinant many-electron
wave functions
\begin{equation}
  \Psi = \sum_i c_i \Phi_i(r_1,\ldots,r_M).
\label{psi}
\end{equation}
$\Psi_i$ are constructed from the single-electron valence basis 
states calculated in the $V^{N-M}$ potential. 
$E$ in (\ref{Schr}) is the valence energy 
(energy needed to remove all valence electrons from the atom).

The effective Hamiltonian has the form
\begin{equation}
  \hat H^{\rm eff} = \sum_{i=1}^M \hat h_{1i} + \sum_{i \neq j}^M \hat h_{2ij} ,
\label{heff}
\end{equation}
$\hat h_1(r_i)$ is the one-electron part of the Hamiltonian
\begin{equation}
  \hat h_1 = c \mathbf{\alpha p} + (\beta -1)mc^2 - \frac{Ze^2}{r} + V^{N-8}
 + \hat \Sigma_1.
\label{h1}
\end{equation}
$\hat \Sigma_1$ is the second order correlation potential which was used for Kr~VIII.

$\hat h_2$ is the two-electron part of the Hamiltonian
\begin{equation}
  \hat h_2 = \frac{e^2}{|\mathbf{r_1 - r_2}|} + \hat \Sigma_2(r_1,r_2),
\label{h2}
\end{equation}
$\hat \Sigma_2$ is the two-electron part of core-valence correlations. It represents 
screening of Coulomb interaction between valence electrons by core electrons.
We also calculate $\hat \Sigma_2$ in the second order of MBPT. The details 
of the calculation of $\hat \Sigma_1$ and $\hat \Sigma_2$ can be found elsewhere
\cite{Kozlov96, Johnson98}. Note however that in contrast to the cited works we now
have no subtraction diagrams.

Only number of electrons changes in the effective Hamiltonian (\ref{heff}) 
when we move from Kr~VII ($M=2$) to Kr~I ($M=8$) while terms $V^{N-8}, \hat \Sigma_1$
and $\hat \Sigma_2$ remain exactly the same.

The results for ground state energy of removal all valence electrons are
compared with experiment in Table~\ref{krypton}. Accuracy of calculations
for all ions and neutral atom are similar and always better than 2\%.

To compare the $V^N$ and $V^{N-8}$ approximations we have also performed
calculations of the ground state energy of Kr~I in $V^N$ potential
with the same size of the basis set and with core-valence correlations
included in the second order of MBPT (including subtraction diagrams).
The result is -18.377~a.u. which differs by only 0.5\% from the result
obtained in $V^{N-8}$ potential and by 1.5\% from the experiment.

\begin{table}
\caption{\label{krypton}Ground state removal energies of Kr~VIII to Kr~I (a.u.).}
\begin{ruledtabular}
\begin{tabular}{lllrr}
 & \multicolumn{2}{c}{State} & Expt\footnotemark[1] & Calc. \\
\hline
Kr~VIII & $4s       $ & $^2S_{1/2}$   &  -4.62317  &  -4.62699 \\
Kr~VII  & $4s^2     $ & $^1S_{0}$     &  -8.70247  &  -8.64060 \\
Kr~VI   & $4s^2 4p  $ & $^2P^o_{1/2}$ & -11.58709  & -11.52481 \\
Kr~V    & $4s^2 4p^2$ & $^3P_{0}$     & -13.96459  & -13.89050 \\
Kr~IV   & $4s^2 4p^3$ & $^4S^o_{3/2}$ & -15.89375  & -15.74736 \\
Kr~III  & $4s^2 4p^4$ & $^3P_{2}$     & -17.25163  & -17.03929 \\
Kr~II   & $4s^2 4p^5$ & $^2P^o_{3/2}$ & -18.14684  & -17.88392 \\
Kr~I    & $4s^2 4p^6$ & $^1S_{0}$     & -18.66132  & -18.28761 \\
\end{tabular}
\end{ruledtabular}
\noindent \footnotetext[1]{NIST, \cite{nist}.}
\end{table}

\subsection{Atoms with two valence electrons}

The fact that $V^{N-2}$ approximation works well for atoms like Mg, Ca, Ba, etc.
is pretty well known. In this section we demonstrate that inclusion of the
higher than second-order core-valence correlations can lead to further significant
improvements in accuracy of atomic calculations. It is much easier to include
higher-order correlations in the $V^{N-2}$ approximation than in any other
potential.

We consider barium atom as an example and start calculations from Ba~II.
Table~\ref{ba+} presents HF and Brueckner energies of Ba~II together with the 
experimental values. Brueckner energies are calculated with the second-order
correlation potential $\hat \Sigma^{(2)}$ and with the all-order
correlation potential $\hat \Sigma^{(\infty)}$. The all-order
$\hat \Sigma^{(\infty)}$ includes screening of Coulomb interaction and
hole-particle interaction (see, e.g. \cite{Dzuba88}). Similar to what
happens for alkali atoms, inclusion of higher-order correlation corrections
for Ba~II reduces the difference between theoretical and experimental
energies from 1 - 2\% to 0.2 - 0.7\%.

\begin{table}
\caption{\label{ba+}Energy levels of Ba~II (cm$^{-1}$).}
\begin{ruledtabular}
\begin{tabular}{lcccc}
\multicolumn{1}{c}{State} & HF & $\hat \Sigma^{(2)}$ & $\hat \Sigma^{(\infty)}$ & 
 Expt\footnotemark[1] \\
\hline
  $6s$       &  75339 & 82318 & 80816 & 80687 \\
  $6p_{1/2}$ &  57265 & 61180 & 60603 & 60425 \\
  $6p_{3/2}$ &  55873 & 59388 & 58879 & 58734 \\
  $5d_{3/2}$ &  68139 & 77224 & 76345 & 75813 \\
  $5d_{5/2}$ &  67665 & 76286 & 75507 & 75012 \\
\end{tabular}
\end{ruledtabular}
\noindent \footnotetext[1]{NIST, \cite{nist}}
\end{table}

Now we are going to use the same correlation potential $\hat \Sigma_1$
for the neutral barium.
The effective Hamiltonian has the form similar to (\ref{heff})
\begin{equation}
  \hat H^{\rm eff} = \hat h_1(r_1) + h_1(r_2) + \hat h_2(r_1,r_2).
\label{bah}
\end{equation}
One-electron part $\hat h_1$ is given by Eq.~(\ref{h1}),
two-electron part $\hat h_2$ is given by Eq.~(\ref{h2}).
For the operator $\hat \Sigma_1$ in (\ref{h1}) we use second-order
correlation potential $\hat \Sigma^{(2)}$ and all-order
correlation potential  $\hat \Sigma^{(\infty)}$, the same as for
the Ba~II ion.

We don't include higher-order correlations in $\hat \Sigma_2$ in present work.
Formally, perturbation expansion for both $\hat \Sigma$-s goes over the same
orders of MBPT. However, calculations show that accurate treatment of
$\hat \Sigma_1$ is usually more important. Since the aim of present work is
to demonstrate the advantages of the $V^{N-M}$ approximation rather than 
presenting best possible calculations, neglecting higher-order correlations
in $\hat \Sigma_2$, which has small effect on final results, is justified.

Table~\ref{barium} shows the results of calculations for few low states 
of Ba~I in the $V^{N-2}$ approximation with $\hat \Sigma^{(2)}$ and
$\hat \Sigma^{(\infty)}$ together with the experimental data.
One can see that inclusion of the higher-order core-valence correlations 
do indeed improve significantly the agreement between theoretical and
experimental data. 

It is interesting to note that there is strong correlation between results
for Ba~I and Ba~II. In both cases the least accurate results are for
states involving $d$-electrons. Inclusion of higher-order core-valence
correlations leads to very similar improvement of results for Ba~II and
Ba~I. Also, if $\hat \Sigma_1$ is rescaled to fit the experimental
energies of Ba~II, the agreement between theory and experiment for Ba~I
would also be almost perfect. This feature can be used to get very accurate
results for negative ions. Experimental results for negative ions are poor
and accurate calculations are difficult. However, if we start calculations
from the $V^{N-M}$ approximation, include $\hat \Sigma$ for core-valence
correlations, rescale $\hat \Sigma_1$ to fit known energies of
a positive ion or neutral atom, the results for a negative ion 
are also going to be very accurate.

\begin{table}
\caption{\label{barium}Two-electron removal energies of Ba~I (a.u.).}

\begin{ruledtabular}
\begin{tabular}{lrccccc}
\multicolumn{2}{c}{State} & Expt\footnotemark[1] & 
$\hat \Sigma^{(2)}$ & $\Delta$(\%)\footnotemark[2] & $\hat \Sigma^{(\infty)}$ 
& $\Delta$(\%)\footnotemark[2] \\
\hline
$6s^2$ & $^1S_0$ &   -0.559152  &  -0.569963  & 1.9  & -0.559032  & 0.02 \\

$6s5d$ & $^3D_1$ &   -0.517990  &  -0.529157  & 2.2  & -0.520645  & 0.67 \\
       & $^3D_2$ &   -0.517163  &  -0.528203  & 2.1  & -0.519799  & 0.51 \\
       & $^3D_3$ &   -0.515423  &  -0.526182  & 2.1  & -0.518029  & 0.51 \\
		  	         	       	      		   
       & $^1D_2$ &   -0.507231  &  -0.516504  & 1.8  & -0.508819  & 0.31 \\
		  	         	       	      		   
$6s6p$ & $^3P_0$ &   -0.503264  &  -0.510328  & 1.4  & -0.502636  & 0.12 \\
       & $^3P_1$ &   -0.501575  &  -0.508609  & 1.4  & -0.500983  & 0.12 \\
       & $^3P_2$ &   -0.497574  &  -0.504472  & 1.4  & -0.497013  & 0.11 \\
		  	         	       	      		   
       & $^1P_1$ &   -0.476863  &  -0.485072  & 1.7  & -0.478031  & 0.24 \\
		  	         	       	      		   
$5d6p$ & $^3F_2$ &   -0.458618  &  -0.466239  & 1.7  & -0.461060  & 0.53 \\
       & $^3F_3$ &   -0.454596  &  -0.461833  & 1.6  & -0.456956  & 0.52 \\ 
       & $^3P_4$ &   -0.450906  &  -0.457765  & 1.5  & -0.453187  & 0.51 \\
\end{tabular}
\end{ruledtabular}
\noindent \footnotetext[1]{NIST, \cite{nist}}
\noindent \footnotetext[2]{$\Delta = |E_{calc}-E_{exp}|/|E_{exp}|\times 100\%$.}
\end{table}

\subsection{Atoms with more than two valence electrons.}

We have demonstrated that $V^{N-M}$ approximation work very well for
atoms with two and eight valence electrons. In is natural to expect
that there are many similar cases in between.

However, there is no reason to believe that this approximation 
works well for all atoms. There are many cases were it doesn't work at all.
It depends mostly on the distances where valence electrons are located
rather than on thier number.
To check whether the $V^{N-M}$ approximation is a good approximation
for a neutral atom it is usually sufficient to perform
Hartree Fock calculations for this atom and check that valence
electrons are localized on larger distances than core electrons.
This is usually the case if valence electrons are in $s$ or $p$ states.
In contrast, $d$ and $f$ valence electrons are localized on distances 
shorter than the distances of the uppermost core $s$ and $p$ electrons.
Their removal would lead to significant change in the atomic core
which means that the $V^{N-M}$ approximation is not good for these atoms.

Roughly speaking, the $V^{N-M}$ approximation should work more or less well
for about half of the periodic table.

\section{Conclusion}

We have demonstrated that the $V^{N-M}$ approximation in which initial
Hartree Fock procedure is done for and ion with all valence electrons
removed, is a good starting point for accurate calculations for
many-electron atoms with $s$ and/or $p$ valence electrons.
The main advantage is relatively simple MBPT for core-valence
correlations which makes it easier to include higher-order
core valence correlations and thus improve the accuracy of the
calculations.

Considering examples of Kr and Ba we have demonstrated that removal
of as many as eight electrons from initial HF potential does not
compromise the accuracy of the calculations for a neutral atom
and that inclusion of the higher-order core-valence correlations do
really lead to significant improvements of the accuracy of the 
calculations.

\bibliography{vn}

\end{document}